\documentclass[a4paper,11pt]{article}
\usepackage{amsmath}
\usepackage{amssymb}
\usepackage{verbatim}
\usepackage{graphicx}
\usepackage{epstopdf}
\DeclareGraphicsExtensions{.eps}
\makeatletter

    \@addtoreset{equation}{section}
\makeatother
\newcommand{\vev}[1]{ \left\langle {#1} \right\rangle }
\newcommand{\cA}{{\cal A}}

\newcommand{\cF}{{\cal F}}

\newcommand{\cO}{{\cal O}}
\newcommand{\cP}{{\cal P}}
\newcommand{\cQ}{{\cal Q}}

\newcommand{\cZ}{{\cal Z}}

\newcommand{\ep}{\varepsilon}

\newcommand\be{\begin{equation}}
\newcommand\ee{\end{equation}}
\newcommand\nn{\nonumber}
\newcommand{\tc}{{\tilde c}}

\newcommand{\fpp}[1]{ \frac{\partial}{\partial {#1}} }

\def\l{\lambda}
\def\v{\varphi}

\author{}
\title{}
\date{}
\begin{document}


\begin{center}
{\LARGE \bf Irregular matrix model with $\mathcal W$ symmetry}
\vskip 12mm
{\large  Sang Kwan Choi\footnote{email:hermit1231@sogang.ac.kr} and Chaiho Rim\footnote{email:rimpine@sogang.ac.kr}}\\
{\it Department of Physics and Center for Quantum Spacetime (CQUeST)}\\
{\it Sogang University, Seoul 121-742, Korea}
\end{center}

\vskip 12mm

\begin{abstract}
We present the irregular matrix model  which has 
$\mathcal{W}_3$  and Virasoro symmetry.
The irregular matrix model is obtained 
using the colliding limit of the Toda field theories
and produces the inner product 
between irregular modules of $\mathcal{W}_3$ symmetry.
We evaluate the partition function 
using the flow equation which is the realization of the  
Virasoro and  $\mathcal{W}$-symmetry. 

\end{abstract}

\vskip 12mm

\setcounter{footnote}{0}

\section{Introduction} 

The irregular matrix model \cite{NR_2012} is obtained 
by the colliding limit of the $\beta$-deformed  
Penner-type matrix model \cite{DV_2009, EM_2009}
which is equivalent to the regular conformal block of primary fields. 
The colliding limit \cite{GT_2012} is the fusion of primary fields at one point 
with their Virasoro momenta infinite
and results in a non-trivial limit. 
Especially, there appears irregular module of rank $n$  which is defined as 
the eigenvector of positive Virasoro generators $L_n, L_{n+1}, \cdots, L_{2n}$.
The irregular module of rank 1 can be constructed 
as the combination of primary and descendant states \cite{G_2009}.
However, for the rank greater than 1  the irregular module is not  simple 
to construct because one needs to take account 
of the consistency condition \cite{CR_2013} 
for the non-negative Virasoro Generators such as $L_0, L_1, \cdots, L_{n-1}$.
The consistency condition is not easy to manipulate algebraically. 
One may detour this difficulty if one uses the 
irregular matrix model and its conformal symmetry. 
In our previous papers, we investigate the case 
with Virasoro symmetry \cite{NR_2012, CR_2013, CRZ_2015, RZ_2015}. 
We may extend this idea to $\mathcal W$ symmetry by considering multi-matrix model.
$\mathcal{W}$ symmetry was previously used to construct 
multi-matrix model of polynomial type in \cite{KMMMP_1992}.
In this paper, we are going to construct the irregular matrix model 
with $\mathcal W$ 
symmetry using colliding limit of the Toda field theory
and present how to find the partition function using the ${\mathcal W}$ symmetry. 

Toda field theory is the generalization of the Liouville field theory and contains 
not only Virasoro symmetry but also higher spin symmetries 
which is summarized in terms of W-symmetry.
Two dimensional Toda field theory associated with simple Lie algebra with rank $k$
is defined by the Lagrangian  
\be
L= \frac1 {8\pi} ( \partial_a \vec \varphi )^2 
+ \mu \sum_{i=1}^k e^{b \,\vec e_i \cdot \vec \varphi }
\ee
where $\vec e_1, \cdots, \vec e_k $ are the simple roots of the Lie algebra.
The bosonic vector field $\vec \varphi $ has $k$ independent components. 
The Toda field theory is conformal provided there is a background charge 
$ \vec Q =  (b + 1/b)  \vec \rho$    
with $\rho$ is the Weyl vector, half of the sum of all positive roots.
In this case, the conformal dimension of the exponential terms is 1 and 
the central charge of the system is $c=k + 12 \vec Q ^2$.

The Toda field theory has $k$ holomorphic symmetry currents 
$W_2, W_3 \cdots W_{k+1}$ 
where $W_2$ is identified as the Virasoro current. 
In this paper we are concentrated on the colliding limit of the $A_2$ Toda field theory 
and construct the $A_2$ irregular matrix model
and find the partition function using the Virasoro and $W_3$ symmetry.
The generalization is straight-forward. 

The paper is organized as follows. Section 2 is devoted to the irregular matrix model.
Starting with $A_2$-Toda field theory, we obtain the irregular matrix model 
using the colliding limit. 
In section 3, we investigate the $\mathcal W$ symmetry in detail. 
The explicit representation of the  $\mathcal W$ symmetry is given as the differential 
operator of the potential variables. 
In section 4, we present how to evaluate the partition function of the irregular matrix model 
only using the symmetries. 
Section 5 is the summary and discussion.

\section{$A_2$ Irregular matrix model}
$A_2$ irregular matrix model is given by two matrix model
and is obtained from the colliding limit of the Penner-type two matrix model. 
For concreteness and comparison,
we will use the  representation as 
appeared in \cite{KMST_2013}.
$A_2$ has two simple roots
which will be represented as  $ \vec e_1= (1, -1, 0)$ and $\vec e_2= (0, 1, -1)$.
The simple roots have the squared length 2. 
The Toda field $\vec \varphi$ is conveniently put as 
$ \vec \varphi=-\frac{\varphi_1}{\sqrt 6} (1,1,-2)-\frac{\varphi_2}{\sqrt 2} (1,-1,0) $
whose components are orthogonal to each other. 
In this convention, the bosonic component satisfies the free field correlation 
$\v_i(z, \bar z) \v_j(w, \bar w) \sim -\delta_{i j} \log |z-w|^2$. 

The vertex primary field $V_a(z_a) =e^{\vec \alpha_a \cdot \vec \varphi (z_a)}$ 
has the conformal dimension 
$\Delta_a= \vec \alpha_a ( \vec Q- \frac12 \vec \alpha_a)$.
We consider the 
$(n+2)$-primary field correlation and 
put the conformal block (holomorphic part of the correlation) 
in terms of the Selberg integral representation 
using the screening operator (exponential terms in the Toda Lagrangian). 
One may  put  the fields, say, one at the infinity $z_{n+1} \to  \infty$, 
one at the origin $z_0 =0 $, and the rest  at  $z_a$ ($a=1, \cdots, n$).
In this case, the conformal block  has the form 
$ \left( \prod_{0 \leq k < \ell \leq n} 
(z_a-z_b)^{-\vec \alpha_a \cdot \vec\alpha_b} \right) \times  Z_{\mathrm{\beta}}$
if one normalizes the result so that the infinite factor $z_{n+1} $ scales away. 
The front multiplicative factor is from the free correlation between the primary fields 
and the rest $Z_{\mathrm{\beta}}$ is due to the correlation between the screenings
 and also with
the primaries;
\be
Z_{\mathrm{\beta}}\equiv \int \prod_{i=1}^{N_1}  d x_i  \prod_{j=1}^{N_2}d y_j~
\Delta(x)^{2 \beta } \Delta(y)^{ 2 \beta } \Delta(x, y)^{-\beta }
e^{ \frac{2b}\hbar \left[
\sum_i V_1 (x_i)+\sum_j V_2 (y_j) \right]}
\ee
where $\Delta(x)=\prod_{i<k} (x_i-x_j)$ and $\Delta(x,y)=\prod_{i,j} (x_i-y_j)$ are 
the Vandermonde determinant which come from the correlation between screening terms. 
We put $\beta =-b^2$ to make the partition function similar to the $\beta$-deformed 
matrix model. 
$N_1$ and $N_2$ are number of screening terms with the root vector $\vec e_1$ and $\vec e_2$ 
respectively. The number  satisfies the neutrality condition \cite{FL_2007}
\be
\sum_{i=0}^n \vec{\alpha}_i+\vec{\alpha}_\infty
+b \sum_{k=1}^2 N_k \vec{e_k}= \vec Q\,.
\ee 

The potential $V_1$  and $ V_2$ appear 
from the correlation between the screening and primary vertex operators.
If one  parametrizes 
$ \vec \alpha_a=\frac{\alpha_a}{\sqrt 3} (1,1,-2)+{\beta_a} (1,-1,0) $,
one has the Penner-type potential   
\be
\frac1\hbar V_1 (x) =- \sum_{a=0}^n \beta_a \log (x-z_a) \,,~~~
\frac1\hbar V_2 (y) =- \sum_{a=0}^n \frac 1 2 (\sqrt3 \alpha_a-\beta_a) \log (y-z_a) \,.
\ee  

The irregular matrix model is obtained if one uses  
the colliding limit 
 similar to the Liouville case \cite{ NR_2012, GT_2012}:
 $\alpha_a , \beta_a \to \infty$ 
and $ z_a \to 0$ so that 
$c_k \equiv \sum_{a=0}^n \alpha_a z_a^{\, k}$ and 
$b_k \equiv \sum_{a=0}^n \beta_a z_a^{\, k}$ 
with $k=0, 1, \cdots, n$  are finite.  
In this case, one has the $A_2$ irregular matrix model ${\cal Z}_n$ 
with rank $n$ whose potential is given in terms of logarithmic and finite 
number of inverse powers 
\be
\frac1\hbar V_1(z)=- b_0 \log z+\sum_{k=1}^n \frac{b_k}{k z^k} \,, ~~~
\frac1\hbar V_2(z)=- \tc_0  \log z+\sum_{k=1}^n  
  \frac{\tc_k}{ k z^k}\,.
\label{potential_V}
\ee  
where we use the short-hand notation $ \tc_\ell = (\sqrt3 c_\ell-b_\ell)/2 $.
In the following, we equally put the irregular matrix model using the new parameter 
$\hbar\equiv -2 i g$ so that  
\be
{\mathcal Z}_{n}
=\int \prod_{i=1}^{N_1} \prod_{j=1}^{N_2} d x_i d y_j
\Delta(x)^{2 \beta} \Delta(y)^{2 \beta} \Delta(x, y)^{-\beta}
e^{-\frac{\sqrt \beta}g \left[
\sum_i V_1 (x_i)+\sum_j V_2 (y_j) \right]} \,.
\label{IRRm}
\ee

\section{$\mathcal{W}$-symmetries of the irregular matrix}
\subsection{Loop equations}
The irregular matrix model \eqref{IRRm}
has two loop-equations. 
One equation is given as the quadratic equation of the resolvent
with its  multipoint defined as
\be
R_{K_1,\cdots,K_s}(z_1;\cdots; z_s)
=\beta\left(\frac g{\sqrt\beta} \right)^{2-s}
\vev{\sum_{i_1=1}^{N_{K_1}} \frac1{z_1-\l_{i_1;K_1}}
\cdots \sum_{i_s=1}^{N_{K_s}} \frac1{z_s-\l_{i_s;K_s}}}_{\mathrm{connected}} \,.
\ee
Denoting $\l_{i;1}=x_i$, $\l_{j;2}=y_j$, one has 
\be
\begin{split}
&R_1(z)^2 +R_2(z)^2- R_1(z) R_2(z)-V_1'(z) R_1(z)-V_2'(z) R_2(z)\\
&+\frac{\hbar Q}2 \left(R_1'(z)+R_2'(z)\right)-\frac{\hbar^2}4 \left(
R_{1;1}(z,z)-R_{1;2}(z,z)+R_{2;2}(z,z) \right)
 = \frac{f_1(z)+f_2(z)}4
 \end{split}
 \label{fullqloop}
\ee
where the quantum correction $f_1, f_2$ are defined as
$f_1(z):=4 g \sqrt{\beta} \sum_i^{N_1} 
\vev{  \frac{-V_1'(z)+V_1'(x_i)}{z-x_i}}$
and 
$f_2(z):=4 g \sqrt{\beta} \sum_i^{N_2} 
\vev{  \frac{-V_2'(z)+V_2'(y_j )}{z-y_j }}$.
Here $\vev{\cdots}$ denotes the expectation value within the matrix integral.
This loop equation is obtained if one performs the conformal transformation 
of the integration variables
$x_i  \to x_i+ \ep/(x_i-z)$
and 
$y_j  \to y_j + \ep/(y_j -z)$.

The other loop equation is given as a cubic equation 
\cite{KLLR_2003, NSW_2003, SW_2009}:
\begin{align}
&0=-R_1^2 R_2+R_1 R_2^2-V_1'(R_1^2+V_1' R_1-\frac{f_1}4)
+V_2'(R_2^2+V_2' R_2-\frac{f_2}4)+\frac{g_1-g_2}4 \nn \\
&+\frac{\hbar Q}4
\left[ 3(V_2' R_2'-V_1' R_1')+R_1 R_2'-R_1' R_2+2(R_2 R_2'-R_1 R_1')
+V_2'' R_2-V_1'' R_1+\frac{f_1'-f_2'}4 \right] \nn \\
&+\frac{\hbar^2 Q^2}8(R_2''-R_1'')
+\frac{\hbar^2}4\left[V_1' R_{1;1}-V_2' R_{2;2}+
R_{1;1} R_2-R_{2;2} R_1-2 R_{1;2}(R_2-R_1) \right] 
 \label{fullcloop} \\
&+\frac{ \hbar^3 Q}{16} \left[R_{1;1}'-R_{2;2}'
+\lim_{\bar z \to z} \left( \fpp{z} R_{1;2} (z , \bar z)
-\fpp{\bar z} R_{1;2}(z, \bar z) \right) \right]
+\frac{\hbar^4}{16}(R_{1;2;2}-R_{1;1;2})\nn \,,
\end{align}
where 
$g_1(z) := 4 g^2 \beta \sum_{i,j}
\vev{\frac{V_1'(x_i)-V_1'(z)}{(z-x_i)(x_i-y_j)}}$
and $g_2(z) := 4 g^2 \beta \sum_{i,j}
\vev{\frac{V_2'(y_j)-V_2'(z)}{(z-y_j)(y_j-x_i)}}$.
This is obtained after varying  the integration variables
$x_i \to x_i+\sum_{j=1}^{N_2} \frac{\epsilon}{(x_i-z)(x_i-y_j)}$
and
$y_j \to y_j+\sum_{i=1}^{N_1} \frac{\epsilon}{(y_j-z)(x_i-y_j)}$.

In this paper, we shall focus on the lowest order of $\hbar$ 
(also putting $Q=0$).
Then, the two equations can be put in a more compact form 
if we define new notations as $R=R_1-R_2/2$, $\tilde{R}=\sqrt 3 R_2/2$,
$\partial \phi_2=V_1'$ and $\partial \phi_1=\frac 1 {\sqrt 3} (2V_2'+V_1')$.
\begin{align} 
&(2R-\partial \phi_2)^2+(2 \tilde{R}-\partial \phi_1)^2 
=-\hbar^2 \xi_2 
\label{loop_W2} 
\\
&(2 \tilde{R}-\partial \phi_1)^3-3 (2 \tilde{R}-\partial \phi_1)(2R-\partial \phi_2)^2
  =-\hbar^3 \xi_3 
\label{loop_W3}
\end{align}
where $\xi_2$ and $\xi_3$ are given explicitly as 
\begin{align} 
 -\hbar^2 \xi_2 &=(\partial \phi_2)^2+(\partial \phi_1)^2+f_1+f_2 
\label{current_W2}
\\
\begin{split} 
 -\hbar^3 \xi_3 
& = -(\partial \phi_1)^3+3(\partial \phi_2)^2 \partial \phi_1
+3(\partial \phi_1-\sqrt{3}\partial \phi_2)(f_1+f_2)  
+3\sqrt{3} \partial \phi_2 f_1
\\
&~~~~-\frac{3}2
(3 \partial \phi_1-\sqrt{3} \partial \phi_2)f_2
-3\sqrt3 (g_1-g_2)\,.
\end{split}
\label{current_W3}
\end{align} 
$\xi_2$ and $\xi_3$ look very complicated but 
turn out to be the representation of  $W_2, W_3$
 currents respectively. This is investigated in the following two subsections. 
 
\subsection{Virasoro current}
One may use  the explicit form of the  potential  \eqref{potential_V}  
to put 
$\partial \phi_1=-\hbar\sum_{k=0}^{n} {c_k}/{z^{k+1}}$,
$\partial \phi_2=-\hbar\sum_{k=0}^{n} {b_k}/{z^{k+1}} $
and  
\be 
f_1(z)+f_2(z)=
- \sum_{k=0}^{n-1} \frac1{z^{k+2} }  
\frac {v_k  (\hbar^2 {\cal Z}_n)} {{\cal Z}_n}
\ee
where we use the translation invariance to put 
$\vev{\sum_i V_1'(x_i)+ \sum_j V_2'(y_j)} =0$. 
In addition, 
the Virasoro differential operator is generalized from the one-matrix model,
which has the form $v_k=v_k^{\partial}$ where
$ v_k^{\partial} \equiv \sum_{\substack{0\leq r \leq n \\ r-s=k}} 
s \left( b_r \fpp{b_s}+c_r \fpp{c_s} \right)  $.
We use the separate notation $ v_k^{\partial}$ for later convenience
since $v_k$ will be modified later for the mode $k<0$.

Therefore,  $\xi_2 $ in \eqref{current_W2}  has the form
\be
\xi_2
=\sum_{k=n}^{2n}\frac{A_k}{ z^{k+2}}
+\sum_{k=0}^{n-1} \frac1{z^{k+2}} 
 \frac { (A_k+v_k^\partial  )   {\cal  Z}_{n}} {{\cal Z}_{n}}  
\ee
where $ A_{k}$  is a constant
\be A_{k}=-\sum_{\ell=0}^k (b_\ell b_{k-\ell}
+ c_\ell c_{k-\ell})\,.
\ee 
We use the convention that $b_\ell=0 = c_\ell$ when $\ell$
does not belong to the element set $\{0, 1 , \cdots, n\}$.
The explicit form $\xi_2$  is identified with the expectation value of Virasoro current
\be
\xi_2=\frac{\vev{\Delta|T(z)|I_n}}{\vev{\Delta|I_n}} \,,
~~~~
T(z)=\sum_{k \in \mathbb{Z}} \frac{L_k}{z^{k+2}} \,,
\ee 
if one recalls that the irregular module  $ |I_n \rangle$  of rank $n$
has the eigenvalue   $  A_k  $ of $L_k$ when  $n \leq k \leq 2n $
and 0 when $k >2n$.   
In addition, this identification 
is consistent with the Virasoro generator representation  
for the parameter set  $\{(b_k, c_k) |~0 \le k \le n\}$
as the differential operator 
${\cal L}_k = A_k +v_k$ 
which is the right action of the irregular module:
$  L_k |I_n \rangle := \mathcal L_k |I_n \rangle $;
\be
\mathcal{L}_k = \left\{ \begin{array}{ll}
0 \,,  &2n <k\\
A_k\,, & n \leq k \leq 2n\\
A_k+v_k\,, & 0 \leq k \leq n-1 \,.
\end{array} \right.
\label{lmode}
\ee

\subsection{$\mathcal W_3$ current}

Let us rewrite $\xi_3$ using the parameters of the potential. 
First note that  
\be
\begin{split}
g_1(z)-g_2(z)=-\hbar \sum_{k=-2}^{n-2}\frac{\hbar^2 b^2}{z^{k+3}}
\vev{ \sum_{i,j}\frac1{x_i-y_j}
 \sum_{r=2}^{n-k}\left( \frac{b_{r+k}}{x_i^r}
+ \frac{\tc_{r+k}}{y_j^r}\right)}\,.
\end{split}
\ee
To put the expectation values as the derivatives of the partition function, 
we use two identities: one is  
\be
4g^2 \beta \vev{\sum_{i,j}\frac{1}{x_i -y_j} 
\frac1{y_j^r}}
=4g^2 \beta \sum_{m=1}^r \vev{
\sum_{j, \ell} \frac1{y_j ^m y_\ell ^{r+1-m}}}
+4 g \sqrt{\beta} \vev{ \sum_j \frac{V_2'(y_j) }{y_j ^{r}}}
\label{identityg2}
\ee
which is obtained if one changes of integration variable  $y_j \to y_j +\ep/y_j^r$
in the partition function ${\cal Z}_n$.
The other is the same as \eqref{identityg2}
with the exchange of $x_i \leftrightarrow  y_j$ and $V_2 \to V_1$
due to the obvious symmetry; 
one may obtain the identity using the integration variable change $x_i \to x_i +\ep/x_i^r$.

Using the above two identities \eqref{identityg2} and its companion, one finds
\be
\begin{split}
g_1(z)-g_2(z)&=
\sum_{k=-2}^{n-2} \frac{\hbar}{z^{k+3}} 
\sum_{r=2}^{n-k}\left[ b_{r+k} \left(\hbar^2 b^2 \sum_{m=1}^{r-1}
\vev{\sum_{i,k} \frac1{x_i^m x_k^{r+1-m}}}-2\hbar^2 b \sum_{p=0}^{n}
\vev{\sum_i \frac{b_{p}}{x_i^{r+p}}} \right) \right. \\
&\left.-\tc_{r+k} \left(\hbar^2 b^2 \sum_{m=1}^{r-1}
\vev{ \sum_{j,\ell}\frac1{y_j^m y_\ell^{r+1-m}}}-2\hbar^2 b \sum_{p=0}^{n}
\vev{\sum_j \frac{\tc_{p}}{y_j^{r+p}}}\right) \right] \,.
\end{split}
\ee
Note that  in the last term there  appear the expectation values of the inverse power  of $y_j$  
up to $(2n+2)$ and the term higher than $n$ is not directly obtained
from the derivative of the  rank $n$ partition function. 
To avoid this difficulty, one may use an extended partition function 
which is partition function with higher rank,
{\it i.e., }  the partition function with the potential with the parameter $c_k, b_k$ 
with $k$ up to $2n+2 $.
In this one has the explicit form of  $\xi_3$ 
\be
\begin{split}
\xi_3&=\sum_{k=2n}^{3n} \frac{ M_k}{ z^{k+3}}
+\sum_{k=n}^{2n-1}\frac1{z^{k+3}}
\frac { ( M_k+\mu_k ) {\cal   Z}_n }
{{  \cal Z}_n} \\
&+\sum_{k=0}^{n-1} \frac1{z^{k+3}} 
\frac{ (M_k+ \mu _k ) {\cal Z}_{2n-k}} 
{{\cal Z}_{2n-k }}
+\sum_{k=-2}^{-1} \frac1{z^{k+3}} 
\frac{ \mu_k ({\cal Z}_{2n-k}) } 
{{\cal Z}_{2n-k }}\Bigg|_{\{b_{k>n},c_{k>n}\} \to 0}
\label{explicitxi_3}
\end{split}
\ee
where $M_k$ 
is a constant
\be M_k= \sum_{r+s+t=k}( 3b_r \, b_s \, c_t-c_r \, c_s\, c_t)\,.
\ee
The terms with inverse powers of $1/z^k$ is carefully rewritten for $-2 \le k \le n-1$ 
by introducing the extended partition functions,
${  \cal Z}_{n+1}, \cdots, {  \cal Z}_{2n+2}$.  
Furthermore, the differential operator $\mu_k$ has the different form 
for the extended case:
\be
\mu_k = \left\{ \begin{array}{ll}
\mu_k^\partial  \,,  & n\le k \le 2n-1 \\
\mu_k^\partial  +\mu_k^{\partial ^2}\,, &-2  \leq k \leq n-1 
\end{array} \right. 
\ee
where 
\be
\begin{split}
\mu_k^{\partial} &\equiv -\sum_{\substack{0\leq r,s \leq n \\r+s-t=k}} \frac t 2
\left(6 b_r c_s \frac{\partial}{\partial b_t}+3 b_r b_s \frac{\partial}{\partial c_t}
-3 c_r c_s \frac{\partial}{\partial c_t} \right) \,, \\
\mu^{\partial^2}_k &\equiv \sum_{\substack{0 \leq r \leq n \\r-s-t=k}} \frac {s \, t}4
\left(6 b_r  \frac{\partial}{\partial b_s}
\frac{\partial}{\partial c_t}+3 c_r \frac{\partial}{\partial b_s} \frac{\partial}{\partial b_t}
-3 c_r \frac{\partial}{\partial c_s} \frac{\partial}{\partial c_t} \right) \,.
\end{split}
\ee

We identify  $ \xi_3$ with  the expectation value of 
the $\mathcal W_3$ current  between 
a regular module $|\Delta \rangle$ 
and an irregular module $|I_n \rangle$ as
\be
\xi_3=\frac{\vev{\Delta|W(z)|I_n}}{\vev{\Delta|I_n}} \,, 
~~~ W(z)=\sum_{k \in \mathbb{Z}} \frac{W_k}{z^{k+3}} \,.
\label{xi3identify}
\ee
Note that  the irregular module has the eigenvalue  $M_k$ of $W_k$  
when $2n \le k \le 3n$. Higher mode $W_k$ with  $k >3n$
annihilates the irregular module.
In this identification one has 
the representation of the $\mathcal W_3$ current 
$ W_k |I_n \rangle := \omega_k |I_n\rangle$ 
where 
\be
\omega_k = \left\{ \begin{array}{ll}
0 \,,  &3n <k\\
M_k\,, & 2n \leq k \leq 3n\\
M_k+\mu_k\,, & n \leq k \leq 2n-1 \\
M_k+\mu_k^{\partial}+\mu_k^{\partial^2}\,, & 0\leq k \leq n-1\,.
\end{array} \right.
\label{omegamode}
\ee
Considering a negative mode acts on the regular module and vanishes
{\it i.e.,} $\vev{\Delta|\omega_k|I_n}=0$ for $k<0$,
one expects the mode with $k=-1,-2$ in  \eqref{explicitxi_3}  to vanish.
This is confirmed in the next subsection.

\subsection{Consistency check of the representation}

One can check the representation
\eqref{lmode} and \eqref{omegamode}
 is compatible with the commutation relation
of the Virasoro and $\mathcal W_3$ modes\cite{FZ_1987}:
\begin{gather}
[L_p, L_q]=(p-q) L_{p+q}+\frac c{12}(p^3-p) 
\delta_{p,-q} \,, \\
[L_p,W_q]=(2p-q)W_{p+q} \,, 
\label{relLW} \\
\begin{split}
-\frac29[W_p,W_q]=&\frac{c}{3 \cdot 5 !}(p^2-1)(p^2-4)p \delta_{p,-q}
+\frac{16}{22+5c}(p-q) \Lambda_{p+q} \\
&+(p-q)\left(\frac1{15}(p+q+2)(p+q+3)
-\frac1 6 (p+2)(q+2) \right) L_{p+q}
\end{split} 
\label{relWW}
\end{gather}
where\footnote{If one rescales
$W_p$ as $i \frac3{\sqrt2} W_p$, then
the algebra reduces to the original
Fateev and Zamolodchikov convention \cite{FZ_1987}.}
\begin{gather}
\Lambda_p=\sum_{k=-\infty}^{\infty} :L_k L_{p-k}:
+\frac15 x_p L_p \,, \nn \\
x_{2\ell}=(\ell+1)(\ell-1) ~~~~~ x_{2\ell+1}=(2+\ell)(1-\ell) \nn
\end{gather}
and the central charge $c=2+12 \vec {Q}^2$.

Note that the irregular module $|I_n \rangle$ with rank $n$ 
is defined as a simultaneous eigenstate of $L_k$'s and $W_k$'s: 
\begin{align}
&L_k |I_n \rangle = A_k |I_n \rangle \,, ~~~~~n \leq k \leq 2n  \\
&W_k |I_n\rangle = M_k |I_n \rangle \,, ~~~ 2n \leq k \leq 3n
\end{align}
where $A_k$ and $M_k$ are eigenvalues. 
In addition, one requires that 
the action of  $L_{k > 2n}$ and $W_{k >3n}$ vanish. 

The eigenvalues are not enough to define the irregular module. 
One needs $L_{k<n}$ and $W_{k<2n}$. 
The Virasoro generator has the representation 
$ {\mathcal L}_k $  for $0 \le k  <n$  in \eqref{lmode} 
and ${\mathcal W}_3$ generator 
$ \omega_k  $ 
for $-2 \le k < 2n$   in \eqref{omegamode}.
However, this representation is not enough 
to check the commutation relation 
\eqref{relWW}.
This is because $\Lambda_p$ contains the negative mode  $L_{k<0}$ of 
Virasoro generators and hence,
the non-negative modes are not closed by themselves in the commutation relation.
On the other hand, one cannot obtain the information of the negative modes 
from the equation \eqref{xi3identify} because the negative mode contribution vanishes 
in the expectation value. Therefore we need another way to find the negative mode
representation.

To find the negative modes 
we consider  
the identity \eqref{identityg2} and its companion
for the extended partition function ${\cal Z}_{n-k}$ for $k<-1$:
\be
\left(v_k^{\partial}+v_k^{\partial^2}\right) {\cal Z}_{n-k}=0 \,,
~~~~v_k^{\partial^2}\equiv -\sum_{-(r+s)=k}
\frac{r \, s}4 \left(\fpp{b_r} \fpp{b_s}+\fpp{c_r} \fpp{c_s} \right)\,.
\label{negativevira}
\ee 
The invariant property of the partition function shows that 
one has  $L_k$ with 
$v_k = v_k^\partial + v_k^{\partial^2}$
for $k<-1$ 
with the extended set of parameters $\{b_{k\geq n}, c_{k\geq0}\}$ when necessary. 

Likewise for ${\mathcal W}_3$, one has the desired identity if 
one considers the transformation  
$x_i \to x_i+\sum_{j}^{N_2}\frac{\epsilon}{(x_i-y_j)x_i^r}$
and $y_j \to y_j+\sum_{i}^{N_1}\frac{\epsilon}{(x_i-y_j)y_j^r}$
to get 
\begin{align} 
& \left(\mu_k^{\partial}+\mu_k^{\partial^2} \right) {\mathcal Z}_{2n-k}=0  ~~{\rm for}~k=-1,-2
\label{k12}
\\ 
& \left(\mu_k^{\partial}+\mu_k^{\partial^2}
+ \mu_k^{\partial^3}\right) {\mathcal Z}_{2n-k}=0  ~~{\rm for}~k\le -3 
\label{k3}
\,, 
\end{align} 
where  
\be 
\label{negativew3}  
\mu_k^{\partial^3}\equiv -\sum_{-(r+s+t)=k}
\frac{r \, s\, t}8 \left(
3\frac{\partial}{\partial b_r} \frac{\partial}{\partial b_s}
\frac{\partial}{\partial b_t}
-\frac{\partial}{\partial c_r} \frac{\partial}{\partial c_s}
\frac{\partial}{\partial c_t} \right) \,.
\ee
Eq \eqref{k12}  
 shows that  one has $\mu_k ({\mathcal Z}_{2n-k})=0$ for $k=-1,-2$,
or $\vev{\Delta|\omega_k|I_n}=0$
as asserted in sec 3.3.
The invariant property of the partition function \eqref{k3}
allows to put 
$\mu = \mu_k^{\partial}+\mu_k^{\partial^2} +\mu_k^{\partial^3}$ 
for $k \leq -3$.

In fact, 
the negative mode representation is easily understood 
if one notes that this representation is found from the coherent state representation 
of Heisenberg algebra
whose positive mode $a_k$ has the eigenvalues $b_k $ or $c_k$
when $k > 0$.
In the coherent state representation, 
the negative mode is  given as the differential operator 
$b_{-k} = -(k/2) \partial /(\partial b_k)$ when $k>0$. 
This is the reason why one and two-derivative terms  appear  
in  the negative Virasoro mode representation 
while one, two and three-derivative terms appear 
in the negative $W$ mode representation as appeared in \cite{KMMMP_1992}.

\section{Irregular partition function}

\subsection{Differential equations}
The loop equations \eqref{loop_W2} and \eqref{loop_W3}
give a series of differential equations for the partition function $\cZ_{n}$.
Large $z$ expansion gives  $2n$-differential equations:
\begin{align}
-v_k \log \cZ_{n}&=d_k ~\,, ~~~ 0 \leq k \leq n-1 \label{differ1}\\
-\mu_k \log \cZ_{n}&= e_k ~\,, ~~~ n \leq k \leq 2n-1 \label{differ2}
\end{align}
where $v_k$ and $\mu_k$ are differential operators defined in
\eqref{lmode}, \eqref{omegamode}.
One may also find differential equations corresponding to $\mu_k$
for $k<n$. However, this equation
contains the extended set of parameters $\{ b_{k>n}, c_{k>n} \}$
and is redundant since we have $2n$-equations \eqref{differ1}, \eqref{differ2}
which will completely fix the partition function.
$d_k$ and $e_k$ in \eqref{differ1} and \eqref{differ2} are given as
\begin{gather}
d_k=A_k+\sum_{r+s=k}\left( d^b_r \, d^b_s+d^c_r \,d^c_s \right) ~\,, ~~~
e_k=M_k+\sum_{r+t+s=k} \left( d_r^c\, d_t^c\, d_s^c-3 d_r^b\, d_t^b\, d_s^c \right)
\label{dkekdef}
\\
d^c_r= \frac{\sqrt3}2 \vev{\sum_j y_j^r}+c_r ~\,, ~~~~~
d_r^b=\vev{\sum_i x_i^r}-\frac{1}2\vev{\sum_j y_j^r}+b_r \,. \nn
\end{gather}
Here we use the convention of the coefficients $b_k=c_k=0$ when $k>n$.
Note that $d_k$ and $e_k$ in  \eqref{dkekdef}
are unknown except $d_0$ 
since they are  given in terms of expectation values.
One has to find these expectation values
as the function of the coefficients of the potential explicitly. 
This can be done by finding the filling fraction
with the resolvents $R_1(z)$ and $R_2(z)$  across a given  branch cut. 
This idea is used to evaluate the partition function
in section \ref{evaluation}.

\subsection{Loop equations and spectral curve}
The resolvents $R_1$ and $R_2$ have the asymmetric role 
in the loop equation. This is because we put the loop equation 
asymmetrically. One may put the loop equation 
in a symmetric way. 
Suppose we introduce 
\begin{align}
u_1(z)&:=R_1(z)+t_1(z) \,,~~~~
t_1(z)=-\frac{2V_1'(z)+V_2'(z)}3 \,, 
\label{defu1}\\
u_2(z)&:=-R_2(z)+t_2(z) \,, ~~~
t_2(z)=\frac{V_1'(z)+2V_2'(z)}3 
\label{defu2}
\end{align}
and $u_0(z):=-u_1(z)-u_2(z)$. 
Then, using  the two loop equations  \eqref{loop_W2} and \eqref{loop_W3}
one finds a spectral curve in a cubic form
\cite{KLLR_2003, NSW_2003, SW_2009, DV_2002}
\be
\prod_{i=0}^2(u -u_i(z))=
u^3+ \frac{\hbar^2\xi_2(z)}4 u - \frac{\hbar^3\xi_3(z)}{12\sqrt3} =0\,.
\label{curve}
\ee
This shows that $u_2$ is the solution of the cubic equation 
which is exactly the same form as the  loop  equation given in  \eqref{loop_W2}
and \eqref{loop_W3}. 
The spectral curve also demonstrates that  
$u_1$ and $u_0$ are two other solutions.
Therefore, $u_1$ and $u_0$ should respect the same loop equation 
\eqref{loop_W2} and \eqref{loop_W3}.

To understand the three branches $u_0, u_1, u_2$, 
we first the case with no quantum correction, {\it i.e.},
 $f_1=f_2=g_1=g_2\equiv 0$.
In this case $\xi_2$ and $\xi_3$ are given in terms of the potentials only
and the three roots $u_1$, $u_2$, $u_0$
are reduced to  $t_1$, $t_2$ and  $t_0(z):=-t_1(z)-t_2(z)$
defined in \eqref{defu1}, \eqref{defu2}.
The classical spectral curve has common points.
For example, when $ t_0(z) =t_1(z)$ 
one finds $ V_1'(z)=0$.
The stationary point of the potential $V_1$ 
corresponds to the double points, 
the intersect of the branches $t_0$ and $t_1$. 
Likewise, $ t_0(z) =t_2(z)$ corresponds to the case 
$ V_2'(z)=0$,
and $ t_1(z)=t_2(z)$ to $ V_1'(z)+V_2'(z)=0$. 
This shows that the three classical  branches 
are connected to each other 
at the stationary points of the potentials.

\begin{figure}
\centering
\includegraphics[width=1\textwidth]{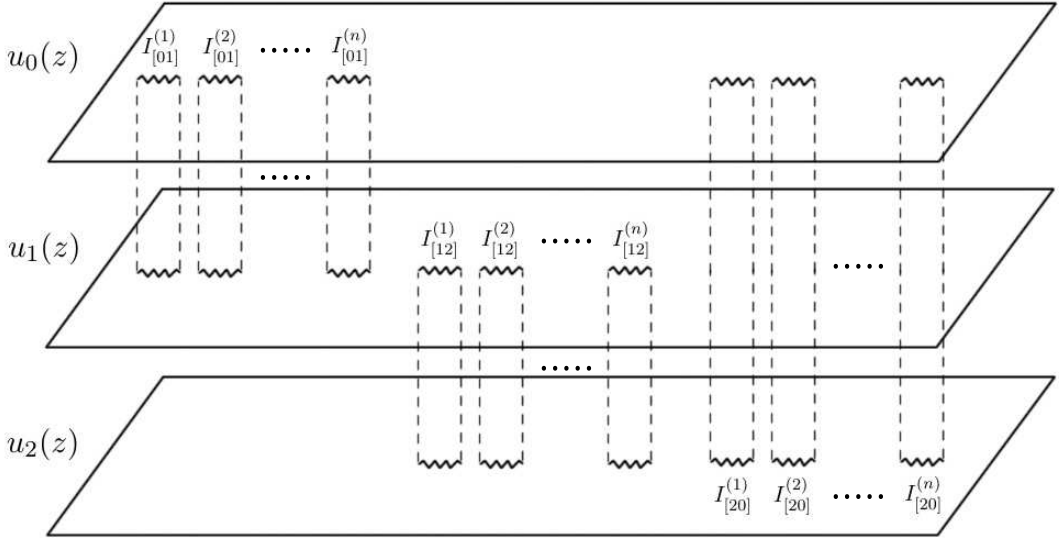}
\caption{Covering space and its cut structure}
\label{f:cut}
\end{figure}

If the spectral curve is deformed by  
$f_1$, $f_2$, $g_1$ and $g_2$, 
then the double point  splits and forms a  branch cut.
As a result,  $3n$-number of branch cuts appear 
 on the complex plane $z$
when  all the double points are distinct. 
Let us denote the branch cuts 
as $I^{(i)}_{[ab]}$ 
connecting two branches $u_a$ and $u_b$ with $i=1, \cdots, n$
(see Fig.\ref{f:cut}).

One may count the number of eigenvalues of the potential 
(number of integration variables) 
by taking the contour integral of the resolvent around the cut. 
Let us denote the contour around the cut $I^{(i)}_{[k,k+1]}$ 
as $\mathcal{A}^{(i)}_{[k,k+1]}$  assuming
the branch index $k \, \rm{mod}\, 3$.
One may find the filling fractions (number of eigenvalues) 
using $u_k$,
\be
\frac{\hbar b}2 n^{(i)}_{k[k+1]} :=
\frac1{2 \pi i}\oint_{\mathcal{A}^{(i)}_{[k,k+1]}} u_k(z) dz 
\ee
or using $u_{k+1}$,
\be
\frac{\hbar b}2 n^{(i)}_{k+1[k]} :=
\frac1{2 \pi i}\oint_{\mathcal{A}^{(i)}_{[k,k+1]}} u_{k+1}(z) dz  \,.
\ee
These two quantities add up to zero; $n_{k[k+1]}^{(i)}+n_{k+1[k]}^{(i)}=0$
since
\be
u_{k}(\l+i0)-u_{k} (\l-i0)=-\bigl(u_{k+1}(\l+i0)-u_{k+1} (\l-i0)  \bigr)
~~~ \mathrm{for} ~~~\l \in I^{(i)}_{[k,k+1]} \,.
\ee
This is due to the fact that $u_k+u_{k+1}+u_{k+2}=0$ and
$u_{k+2}$ is continuous on the cut $I^{(i)}_{[k,k+1]}$.
Therefore, one may have $N=N_1+N_2$ where
\be
N_1=\sum_{i=1}^n \left(n^{(i)}_{1[0]}+n^{(i)}_{1[2]}\right)\,, ~~~
N_2=-\sum_{i=1}^n \left(n^{(i)}_{2[1]}+n^{(i)}_{2[0]} \right) \,.
\label{evcon}
\ee
(Note the minus sign in $N_2$
comes from the definition of $u_2$ in \eqref{defu2}.)

\subsection{Evaluation of partition function}
\label{evaluation}
We evaluate the partition function of the rank $1$ explicitly
in this subsection.
The rank $1$ partition function has two differential equations
\be
-v_0 \log \cZ_{1}=d_0 \,,~~~~
-\mu_1 \log \cZ_{1}=e_1
\label{diffeq}
\ee
where $v_0=b_1 \fpp{b_1}+c_1 \fpp{c_1}$,
$\mu_1=-3 b_1 c_1 \fpp{b_1}-\frac32 (b_1^2-c_1^2) \fpp{c_1}$
and $d_0$, $e_1$ are defined in \eqref{dkekdef}.
Since $d_0$ is a constant
\be
d_0=(b N_1)^2-b N_1 \, b  N_2+(b N_2)^2+2 b_0 \, b N_1
-b_0 \, b N_2+\sqrt3 c_0 \, b N_2
\ee
one has the solution of the first equation in \eqref{diffeq} of the form,
\be
\log \cZ_{1}=-d_0 \log c_1+H(t)
\ee
where $t:=b_1/c_1$ and  $H(t)$ is a homogeneous solution to $v_0$.

Plugging this into the second one in \eqref{diffeq}, one obtains
\be
(3-t^2)t \frac{\partial H(t)}{\partial t}
=\frac23\frac{e_1}{c_1}+(t^2-1)d_0 \,.
\label{tdiffeq}
\ee
where $e_1$ is given in terms of the expectation values 
$\vev{\sum_i x_i}$ and $\langle \sum_j y_j \rangle$,
\be
\begin{split}
e_1=&3b_0^2 c_1-3c_0^2 c_1+6 b_0 c_0 b_1
+\frac38 \left(2c_0+\sqrt3 N_2\right)^2
\Bigl(2c_1+\sqrt3 \big\langle \sum_j y_j \big\rangle \Bigr) \\
&-\frac38\left(2b_0+2N_1-N_2\right)
\biggl[2\Bigl(2b_1+2\big\langle \sum_i x_i \big\rangle- \big\langle \sum_j y_j
 \big\rangle\Bigr)
\left(2c_0+\sqrt3 N_2 \right) \\
& +\left(2b_0+2N_1-N_2 \right)
\Bigl(2c_1+\sqrt3 \big\langle \sum_j y_j \big\rangle \Bigr) \biggr] \,.
\end{split}
\ee
To solve the equation \eqref{tdiffeq},
we need the explicit from of $e_1/c_1$
as the function of $t$.
This can be done using the constraint of the filling fraction.

For the rank $1$, we have three cuts $I_{[01]}$, $I_{[12]}$ and $I_{[20]}$
whose classical double points are $-\frac{b_1}{b_0}$,
$\frac{-b_1-\sqrt3 c_1}{b_0+\sqrt3 c_0}$
and $\frac{-b_1+\sqrt3 c_1}{b_0-\sqrt3 c_0}$, respectively.
However,   one filling fraction is enough, say
\be
\frac1{2 \pi i} \oint_{\cA_{[01]}} u_1(z) dz=\frac{\hbar b}2n_{1[0]} \,.
\label{cint}
\ee

The branch $u_1$ in \eqref{curve} is given as
\be
\frac1{\hbar}u_1(z)=
\frac{\left(\sqrt3+i 3\right) \cP_2+
\left(\sqrt3-i 3\right)\left(\cP_3+\sqrt{\cP_3^2-\cP_2^3}\right)^{2/3}}
{12\left(\cP_3+\sqrt{\cP_3^2-\cP_2^3}\right)^{1/3}}
\label{u1int}
\ee
where $\cP_2$ and $\cP_3$ are polynomials given by
\begin{gather}
\cP_2(z)=(b_0^2+c_0^2+d_0) z^2+2(b_0 b_1+c_0 c_1) z+b_1^2+c_1^2 \,, \nn\\
\begin{split}
\cP_3(z)=&(c_0^3-3 b_0^2 c_0+e_0) z^3
+(3c_0^2 c_1-3 b_0^2 c_1-6 b_0 c_0 b_1+e_1) z^2 \\
&-3(c_0 b_1^2+2 b_0 b_1 c_1-c_0 c_1^2)z+c_1^3-3 b_1^2 c_1\,. \nn
\end{split}
\end{gather}
Here $e_0$ is a constant, 
$e_0=3b_0^2 c_0-c_0^3-3(N_1-\frac{N_2}2+b_0)^2
(\frac{\sqrt3}2 N_2+c_0)+(\frac{\sqrt3}2 N_2+c_0)^3$.
Six roots of $(\cP_3^2-\cP_2^3)$
implies three branch cuts on $z$-plane.
(Note that a cubic branch cut whose branch points are
roots of $\left(\cP_3+\sqrt{\cP_3^2-\cP_2^3}\right)$
is not reduced to the classical point 
and no eigenvalues lie on it. Thus we don't need to take account of this cut.)

To find the filling fraction perturbatively,
we assume that $|t| \ll 1$.
In this case, the cut $I_{[01]}$
is widely separated from the others:
Rescaling the integration variable 
in the contour integral \eqref{cint} with $b_1$,
the cut $I_{[01]}$ has a finite width whereas
other cuts shrink to a point as $t \to 0$.
Thus, one can expand the $u_1(z)$ in  \eqref{u1int} safely in powers of $t$
to get 
$\left(\cP_2^3-\cP_3^2\right) =b_1^2 c_1^4
\left[Q_0+ t Q_1+ \cO(t^2) \right]$.
However, 
the explicit form of $Q_i$ depends on the relative scales of $e_1$.
Small $t$-expansion assumes that 
$|b_1|< |c_1| $. In addition, 
since $e_1$ is a quantum deformation, 
$|c_1|<|e_1|$ is not allowed.
Therefore, we have two different relative scales:  
(I) $|e_1|\lesssim|b_1| \ll |c_1|$ and
(II) $|b_1| \ll |e_1| \lesssim|c_1|$.

Let us first consider the case  (I).
In this case $\tilde{e}_1:=e_1/b_1$ is small.
Assuming  $\tilde{e}_1 =\cO(t^0)$ at most,
one has 
\begin{align}
Q_0(z)&=3 \left((3b_0^2+d_0)z^2+6 b_0 z+3\right)
\nn\\ 
Q_1(z)&=z\left((36b_0^2 c_0+12 c_0 d_0-2 e_0) z^2
+(72 b_0 c_0 -2 \tilde{e}_1)z+36 c_0 \right)\,.
\end{align}
The filling fraction $n_{1[0]}$ is given  in power of t,
\be
\begin{split}
b n_{1[0]} &=\frac1{\pi i}\oint_{\cA_{[01]}} dz \left[
\frac{ c_0 z+ \sqrt{\cQ_0(z)/3}}{2\sqrt3 z^2}
-t \frac{\tilde{e}_1+e_0 z-d_0 \sqrt{\cQ_0(z)/3}}
{6 \sqrt3 \sqrt{\cQ_0(z)/3}}+\cO(t^2) \right] \\
&=-b_0+\sqrt{b_0^2+\frac{d_0}3}
-t \frac{(d_0+3 b_0^2) \tilde{e}_1-3 b_0 e_0}{3 \sqrt3 (d_0+3 b_0^2)^{3/2}}
+\cO (t^2) \,.
\end{split}
\ee 
Noting $\tilde{e}_1 =\cO(t^0)$,
one has to require the filling fraction 
$b n_{1[0]}= -b_0+\sqrt{b_0^2+{d_0}/3} + t b \Delta  n_{1[0]}$
so that $\Delta  n_{1[0]} = \cO(t^0)$.
Since we regard $ n_{1[0]}$ as the controlling parameter,
we force $\Delta  n_{1[0]}=0$ so that 
\be
b n_{1[0]}= -b_0+\sqrt{b_0^2+{d_0}/3} \,.
\ee
In this case, one has 
\be
\tilde{e}_1=\frac{3 b_0 e_0}{3 b_0^2+d_0}
-t \frac{3\left(d_0(d_0^3+2c_0 d_0 e_0-e_0^2)
+b_0^2(3d_0^3+6c_0 d_0 e_0+ 2e_0^2)\right)}
{4 (3 b_0^2+d_0)^3}+\cO(t^2)  
\label{tildee1}
\ee
and finds the partition function from \eqref{tdiffeq}
\be
\cZ_{1}=\mathcal{N}\, b_1^{-\frac{d_0}3} c_1^{-\frac{2 }3 d_0}
\, e^{\frac{2b_0 e_0}{9 b_0^2+3d_0}t +\cO(t^2)}
\label{resb1>e1}
\ee
where $\mathcal{N}$
is the normalization independent of $t$.

For the case (II), one finds a quite different structure from that of (I).
Note that $|b_1| < |e_1|$.
This parameter domain allows to put $b_1=0$.
In addition, 
$b_1$ should be zero when the quantum correction $e_1$ is zero.
In this case, 
the double point of the classical branches of $t_0$ and $t_1$
does not exist on the finite domain of the complex plane
since it is now given by the stationary point of $V_1$, satisfying $V_1'(z) =\frac{b_0}z=0$.
Considering this, one may force $ n_{1[0]}=0$ which is true at the classical level.
With $b_1$ which can be 0, we had better introduce a new parameter 
$\hat{e}_1:=e_1/c_1$ instead of $\tilde{e}_1$.
Assuming $\hat{e}_1 =\cO(t^0)$ at most,
one has
\begin{align}
\cQ_0(z)&=(9 b_0^2+3 d_0-2 \hat e_1)z^2+18 b_0 z+9
\nn\\
\cQ_1(z)&=z \left((36 b_0^2 c_0+12 c_0 d_0-2 e_0-6 c_0 \hat e_1) z^2
+72 b_0 c_0 z+36 c_0 \right)
\end{align} and the filling fraction has the form
\be
\begin{split}
bn_{1[0]}&=\frac1{\pi i} \oint_{\cA_{[01]}}
dz \left[ \frac{\sqrt3 c_0 z+ \sqrt{\cQ_0(z)}}{6 z^2} 
+t\frac{9(c_0 \hat{e}_1-e_0)z+\sqrt3(3 d_0+\hat{e}_1)
\sqrt{\cQ_0(z)}}{54 \sqrt{\cQ_0(z)}}
+\cO(t^2)\right]\\
&=-b_0+\frac{\sqrt{9 b_0^2+3 d_0-2 \hat{e}_1}}3
+t \frac{b_0 (e_0-c_0 \hat{e}_1)}{(9 b_0^2+3d_0-2 \hat{e}_1)^{3/2}}+\cO(t^2) \,.
\label{bn0}
\end{split}
\ee
Putting $n_{1[0]}=0$, one finds
\be
\hat e_1=\frac32 d_0+t \frac{2 e_0-3 c_0 d_0}{2 b_0} +\cO(t^2)
\ee
and the partition function  from \eqref{tdiffeq},
\be
\cZ_{1}=\hat{\mathcal{N}}c_1^{-d_0}
\, e^{\frac{2e_0-3c_0 d_0}{9 b_0}t +\cO(t^2)}
\label{resb1<e1}
\ee
with  $\hat{\mathcal{N}}$ a normalization independent of $t$.
In this evaluation, we assume that $b_1>0$ and the vanishing filling fraction 
is obtained when $b_0>0$. The same is true if one assume $b_1<0 $ with $b_0<0$.
The vanishing filling fraction condition holds as far as $b_0 b_1 >0$,
which is the case $V_1$ having no stationary point in our complex domain. 
(Note that due to the logarithmic potential, we exclude the negative real axis in our complex domain).

Suppose we put $b_1 \equiv 0$ from the beginning
and require the filling fraction $n_{1[0]}$ to vanish.
In this case, one finds that 
the  partition function 
is simply  reduced to $\cZ_{1}=\hat{\mathcal{N}}c_1^{-d_0}$
since  $t=0$.
In addition,  the two differential operators $v_0$ and $\mu_1$
are not independent but has the relation $\mu_1=\frac32 c_1 v_0$. 
The case with $b_1=0$ is identified in \cite{KMST_2013} with the semi-degenerate 
where the regular module has the null vector at the first level
\cite{FZ_1987, W_2009, KMST_2010}.
In our approach, if the semi-degenerate module
is put at infinity, we have the semi-degenerate $\langle \Delta|$ in \eqref{xi3identify}
whose conformal dimension is $\Delta=\vec \alpha_\infty (\vec Q-\frac12
\vec \alpha_\infty)$ with $\vec \alpha_\infty= \varkappa \vec w_2$
where $\vec w_2$ is the fundamental weight satisfies
$\vec w_i \cdot \vec e_j=\delta_{ij}$.
In this case, we have $N_1=0$ as the semi-degenerate result.

\section{Summary and discussion}
We generalize the (Virasoro) irregular matrix model 
so that the model contains the  Virasoro and $\mathcal{W}_3$ symmetry.
This model is constructed 
using the colliding limit of $A_2$ Toda field theory.  
The symmetry of the irregular models 
is analyzed through the loop equations 
which have the quadratic and cubic form. 
The spectral curve obtained
corresponds to the Seiberg-Witten curve of the $SU(3)$ 
super-conformal linear quiver theory.

It should be emphasized the spectral curve \eqref{curve}
is enough to find the partition function 
of the irregular matrix model 
without evaluation of the functional integral or Selberg integral.
Using the  differential representation of the 
Virasoro and $\mathcal{W}_3$ symmetry generators
we derive  the differential  equations for the partition function 
from the loop equations.
The differential equations allow us to find the partition function of the 
irregular model. 
We present the explicit form of the representation 
and find the partition function to the lowest order of $\hbar$ 
(corresponding to the large $N$ limit) for the non-trivial case 
(irregular module with rank 1). 
It is not  difficult to find the partition function with   rank greater than 1 
if one uses the parameter scale as  $\left| \frac{b_{k+1}}{b_k}\right| 
\ll \left|\frac{b_k}{b_{k-1}}\right|$,
$\left| \frac{c_{k+1}}{c_k}\right| \ll \left|\frac{c_k}{c_{k-1}}\right|$ 
and $\left|\frac{b_1}{b_0}\right| \ll \left|\frac{c_n}{c_{n-1}}\right|$
as was used in Liouville case \cite{NR_2012, CR_2013}.

Once the partition function is known, 
one may construct the irregular conformal block(ICB),
noting that the partition function with appropriate potential is related 
with an inner product between an irregular module and regular/irregular modules. 
The simplest ICB, the inner product $\vev{I_m|I_n}$
is given  in terms of irregular matrix model.  
One may consider the irregular partition function $\cZ_{(m:n)}$ 
where irregular module of rank $n$ lies at the origin and 
one with rank $m$ at infinity: 
\begin{gather}
\cZ_{(m:n)}
=\int \prod_{i=1}^{N_1} \prod_{j=1}^{N_2} d x_i d y_j
\Delta(x)^{2 \beta} \Delta(y)^{2 \beta} \Delta(x, y)^{-\beta}
e^{-\frac{\sqrt \beta}g \left[
\sum_i V_1^{(m:n)} (x_i)+\sum_j V_2^{(m:n)} (y_j) \right]} \,, \\
\frac{V_1(z)}\hbar=- b_0 \log z+\sum_{k=1}^n \frac{b_k}{k z^k}
+\sum_{\ell=1}^m \frac{b_{-\ell}z^\ell}{\ell} \,, ~~~
\frac{V_2(z)}\hbar =- \tc_0  \log z+\sum_{k=1}^n  
  \frac{\tc_k}{ k z^k }+\sum_{\ell=1}^m \frac{\tc_{-\ell}z^\ell}{\ell} \,. \nn
\end{gather}
There are a few subtle issues when one identifies 
this partition function with the inner product 
$\vev{I_m|I_n}$.
One is the extra contribution due to the colliding limit:
$\prod_{a,b}(1-z_b/\bar z_a)^{-\vec \alpha_a \cdot \vec \alpha_b}
\to e^{\zeta_{(m:n)}}$ with 
$\zeta_{(m:n)}=\sum_k^{min(m,n)}2(b_k b_{-k}+c_k c_{-k})/k$
as $z_b \to 0$, $\bar z_a \to \infty$.
This non-trivial contribution  was considered in \cite{CRZ_2015}
for $A_1$ case.
The other is about the normalization for the case with rank greater 1.
One has to take account the normalization properly
because of the consistency condition for the Virasoro and $\mathcal W$ symmetry
\cite{CR_2013}.
This consideration leads to the irregular conformal block $\cF^{(m:n)}$;
\be
\cF^{(m:n)}=
\frac{e^{\zeta_{(m:n)}}\cZ_{(m:n)}}
{\cZ_{(0:n)} \cZ_{(m:0)}} \,.
\ee

One may find the complete representation
of the symmetry generators  which has
non-leading order of $\hbar$ with  $Q \neq 0$ without difficulty
considering the full equations given in \eqref{fullqloop}, \eqref{fullcloop}.
In addition, the irregular matrix model is easy to generalize into $A_k$ model 
since it is composed of  $k$-matrix potential along with the 
Vandermonde determinant 
whose power is given in terms of Dynkin index. 
There will appear more than cubic power in the loop equations.
To control these complicated equations, 
one may resort to DDAHA as demonstrated in \cite{MRZ_2014}.
This will be  presented elsewhere in the future. 

Finally, the irregular matrix model is motivated by Argyres-Douglas theory \cite{AD_1995,APSW_1996}
in connection with AGT conjecture \cite{AGT}, 
which develops irregular punctures to the holomorphic one form of the Hitchin system \cite{GMN_2009,BT_2009,NX_2009}.
So far, the Virasoro case has been intensively investigated using the irregular matrix model. 
Extension to $\mathcal W$ symmetry will provide a useful tool to study Argyres-Douglas theory
corresponding to $SU(N)$ gauge theory.

\subsection*{Acknowledgement}
This work was supported by the National Research Foundation of Korea(NRF) 
grant funded by the Korea government(MSIP) (NRF-2014R1A2A2A01004951).

\end{document}